\documentstyle[12pt]{article}

\newcommand {\vs}[1]  { \vspace*{#1 cm} }

\newcounter{eq}
\newcounter{sc}

\newcommand {\MPL}  {Mod. Phys. Lett.}
\newcommand {\IJMP}  {Int. J. Mod. Phys.}
\newcommand {\NP}   {Nucl. Phys.}

\newcommand {\PL}   {Phys. Lett.}

\newcommand {\PR}   {Phys. Rev.}
\newcommand {\PRL}   {Phys. Rev. Lett.}

\newcommand {\AP}   {Ann. of Phys.}
\newcommand {\PTP}  {Prog. Theor. Phys.}

\newcommand {\CQG}  {Class. Quantum. Grav.}

\def\overleftrightarrow#1{\vbox{\ialign{##\crcr
 $\leftrightarrow$\crcr\noalign{\kern-1pt\nointerlineskip}
 $\hfil\displaystyle{#1}\hfil$\crcr}}}

\newlength{\minitwocolumn}
\begin{document}

\begin{flushright}
DPUR/TH/21\\
April, 2010\\
\end{flushright}
\vspace{30pt}
\pagestyle{empty}
\baselineskip15pt

\begin{center}
{\large\bf Remarks on Higgs Mechanism for Gravitons

 \vskip 1mm
}

\vspace{10mm}

Ichiro Oda
          \footnote{
           E-mail address:\ ioda@phys.u-ryukyu.ac.jp
                  }

\vspace{10mm}
          Department of Physics, Faculty of Science, University of the 
           Ryukyus,\\
           Nishihara, Okinawa 903-0213, JAPAN \\

\end{center}


\vspace{10mm}
\begin{abstract}
We construct two kinds of model exhibiting Higgs mechanism for gravitons in 
potentials of scalar fields. One class of the model is based on a potential 
which is a generic function of the induced internal metric $H^{AB}$, and the
other involves a potential which is a generic function of the usual metric tensor 
$g_{\mu\nu}$ and the induced curved metric $Y_{\mu\nu}$. In the both models, 
we derive conditions on the scalar potential in such a way that gravitons 
acquire mass in a flat Minkowski space-time without non-unitary propagating modes 
in the process of spontaneous symmetry breaking of diffeomorphisms through 
the condensation of scalar fields. We solve the conditions and find a general 
solution for the potential. As an interesting specific solution, we present 
a simple potential for which the Higgs mechanism for gravitons holds in any value
of cosmological constant. 
\vspace{15mm}

\end{abstract}

\newpage
\pagestyle{plain}
\pagenumbering{arabic}


\rm
\section{Introduction}

Since the advent of the pioneering work of Fierz and Pauli \cite{Fierz},
it has turned out that the theoretical problems involved for constructing
a complete massive gravity theory are very subtle, challenging and even call
for consideration beyond perturbation theory \cite{Vainshtein}.

For instance, it is nowadays well known that there is no smooth massless
limit in perturbation theory of massive gravity in the sense that the
massless limit in massive gravity $\it{does}$ exist but does $\it{not}$ agree with the 
result predicted by Einstein's general relativity which describes massless 
gravitons. This mass discontinuity, which is dubbed as van Dam-Veltman-Zakharov 
(vDVZ) discontinuity \cite{Veltman, Zakharov}, is a fatal defect in massive gravity theory 
since it produces a different value for the bending of light by the sun, 
that is, the value obtained from massive gravity is $\frac{3}{4}$ of that from 
Einstein's general relativity and experiments,
which implies that gravitons must be strictly massless in nature.

The reason why there is no smooth massless limit is that there is a discrepancy
of the number of dynamical degrees of freedom between massive and massless 
gravitational theories. Actually, in four space-time dimensions, 
we have five states of spins $\pm 2$, $\pm 1$ and $0$ for massive gravitons 
whereas we have only two states of helicity $\pm 2$ for massless gravitons, so 
in the massless limit of massive gravitons, an extra helicity $0$ state 
shows up in the spectrum, thereby breaking a smooth limit to massless ones. 
(Two states of spins $\pm 1$ decouples smoothly because of the current conservation.)
The same problem is known to exist in non-abelian gauge theories though
there is no such a problem in case of abelian gauge theories by the current 
conservation \cite{Vainshtein}.

It is worthwhile to notice that this observation suggests us one resolution for 
the vDVZ discontinuity, namely a possible way out of this might be to match the number 
of the degrees of freedom in the both theories. 
Indeed, in case of massive non-abelian gauge theories, we can take a smooth massless limit 
by incorporating extra scalar fields and triggering vacuum condensation of them, 
which is nothing but the Higgs mechanism.  
Thus, it is very natural to pursue the analogy of massive gauge theories and then
ask ourselves whether it is possible to construct the Higgs mechanism for gravitons
in order to obviate the problem relevant to the absence of massless limit in
massive gravity theories.

Recently, interests on the construction of massive gravity theories have
revived from different physical motivations \cite{Percacci}-\cite{Rubakov2}. 
One motivation comes from the astonishing 
observational fact that our universe is not just expanding but is at present 
in an epoch of undergoing an accelerating expansion \cite{Riess, Perlmutter, WMAP}. 
Massive gravity theories might shed some light on this problem in the sense that 
they could modify Einstein's general relativity at large cosmological scales 
and might lead to the present accelerated expansion of the universe without assuming 
the existence of mysterious dark matter and dark energy. 

The other motivation for attempting to construct massive gravity theories
is conceptual and is related to the noncritical string theory applied to 
quantum chromodynamics (QCD) \cite{'t Hooft}.
For instance, as inspired in AdS/CFT duality, if we wish to apply a bosonic string theory 
to the gluonic sector in QCD, massless fields such as spin 2 graviton
in string theory, must either become massive or be removed somehow by an ingenious 
dynamical mechanism since such the massless fields do not appear in QCD.

A few years ago, 't Hooft proposed a new Higgs mechanism for gravitons where 
the massless gravitons '$\it{eat}$' four real scalar fields and consequently
become massive \cite{'t Hooft}. In his model, vacuum expectation values (VEV's) 
of the scalar fields are taken to be the four space-time coordinates by gauge-fixing 
diffeomorphisms, so the whole diffeomorphisms are broken spontaneously. 
Afterward, a topological term was added to the 't Hooft model 
where an '$\it{alternative}$' metric tensor is naturally derived and the topological 
meaning of the gauge conditions was clarified \cite{Oda} \footnote{Similar but 
different approaches have been already taken into consideration in Ref. \cite{Oda1}.}.  

One serious drawback in the 't Hooft model is that a scalar field appearing 
after the SSB is a non-unitary propagating field so that in order to keep the
unitarity the non-unitary mode must be removed from the physical Hibert space 
in terms of some procedure. This problem was solved by including higher-derivative 
terms in scalar fields and tuning appropriately the cosmological constant 
to be a negative value in Ref. \cite{Kaku2} \footnote{A different formalism was 
constructed in Ref. \cite{Demir}.}. 

More recently, Chamseddine and Mukhanov have presented a new Higgs mechanism for
gravitons also by adding a specific form of higher-derivative terms in scalar fields to 
the Einstein-Hilbert action \cite{Chamseddine}. One advantage of their model is 
that we do not have to
restrict the cosmological constant to be negative, namely zero or positive cosmological
constant is also allowed to trigger the gravitational Higgs mechanism. This model was
later examined in \cite{Oda2} from the viewpoint of general models with an arbitrary
potential of the induced internal metric. 

The aims of this article are the following: First, we construct two kinds of massive gravity
models based on either the induced internal metric $H^{AB}$ or the usual metric tensor 
$g_{\mu\nu}$ and the induced curved metric $Y_{\mu\nu}$. Although it seems that there are 
two distinct formulations of massive gravity models, the two formulations are in essence
equivalent through identities of the metrics. Next, we derive conditions on 
the scalar potential in such a way that gravitons acquire mass in a flat Minkowski 
space-time without non-unitary propagating modes in the process of spontaneous 
symmetry breaking of diffeomorphisms. We solve the conditions and find a general solution
for the potential. As an interesting specific solution, we present a simple potential 
for which the Higgs mechanism for gravitons holds in any value of the cosmological constant. 

This paper is organized as follows: In section 2, we construct a model of the Higgs
mechanism for gravitons by using the induced internal metric $H^{AB}$.  In section 3, 
we also construct such a model by using the usual metric tensor $g_{\mu\nu}$ and the induced 
curved metric $Y_{\mu\nu}$.
In section 4, we present some concrete models satisfying the conditions derived so far.
In particular, we present a simple model showing the Higgs mechanism 
for gravitons in any value of the cosmological constant in four dimensions.
The final section is devoted to conclusions and discussion.

\section{Model based on the induced internal metric $H^{AB}$}

In this section, we wish to construct a rather general model of Higgs mechanism for 
gravitons on the basis of the induced internal metric $H^{AB}$ in a general $D$-dimensional 
space-time.
We start with the following action \cite{Oda2} \footnote{We obey the conventions 
and the notation in the Misner et al.'s textbook \cite{MTW}.}:
\begin{eqnarray}
S = \frac{1}{16 \pi G} \int d^D x \sqrt{-g} [ R - V(H^{AB}) ].
\label{Model 1}
\end{eqnarray}
Here $G$ is the $D$-dimensional Newton's constant and the induced internal metric 
$H^{AB}$ is defined as
\begin{eqnarray}
H^{AB} = g^{\mu\nu} \nabla_\mu \phi^A \nabla_\nu \phi^B,
\label{H}
\end{eqnarray}
where $\phi^A$ are $D$ real scalar fields with $A = 0, \cdots, D-1$, and
the indices $A, B, \cdots$ are raised and lowered in terms of
the Minkowski metric $\eta_{AB} = diag(-1, 1, \cdots, 1)$ \footnote{Of course,
we could also consider a curved metric $g_{AB}$ for the internal space 
instead of this flat metric $\eta_{AB}$ at the expense of simplicity of
the formalism.}.
Finally, $\it{a \ priori}$ $V$ is a generic function of $H^{AB}$.

This general action gives us the following equations of motion:
\begin{eqnarray}
\nabla^\mu ( \frac{\partial V}{\partial H^{AB}} \nabla_\mu \phi^B ) &=& 0,
\nonumber\\
G_{\mu\nu} \equiv R_{\mu\nu} - \frac{1}{2} g_{\mu\nu} R 
&=& - \frac{1}{2} g_{\mu\nu} V + \nabla_\mu \phi^A \nabla_\nu \phi^B 
\frac{\partial V}{\partial H^{AB}}.
\label{Eq. of motion 1}
\end{eqnarray}

We are now interested in obtaining 'vacuum' solution of the form
\begin{eqnarray}
\phi^A &=& x^\mu \delta_\mu^A, 
\nonumber\\
g_{\mu\nu} &=& \eta_{\mu\nu}.
\label{Vacuum}
\end{eqnarray}
This vacuum solution is not static since one component of $\phi^A$, that is,
$\phi^0$ is essentially equivalent to time $x^0 = t$.
The requirement of the presence of the vacuum solution 
leads to a constraint on the potential $V$
\begin{eqnarray}
\frac{\partial V(H_*)}{\partial H^{AB}} = \frac{1}{2} \eta_{AB} V(H_*),
\label{Condition 1}
\end{eqnarray}
where we have defined $H^{AB}_* = \eta^{AB}$ and omitted to write 
the indices $AB$ on $H^{AB}_*$ explicitly for simplicity. In other words,
the equation (\ref{Condition 1}) is a constraint imposed on the potential $V$ in order to
have a flat Minkowski space-time as the background. 

Next, we expand the fields around this vacuum (\ref{Vacuum}) as 
\begin{eqnarray}
\phi^A &=& x^\mu \delta_\mu^A + \varphi^A, 
\nonumber\\
g_{\mu\nu} &=& \eta_{\mu\nu} + h_{\mu\nu},
\label{Fluctuation}
\end{eqnarray}
and write out all the terms up to second order.

At this stage, we gauge away the scalar fluctuations $\varphi^A$ by using diffeomorphisms. 
Of course, once we gauge away the $D$ scalars, we can no longer gauge away any components 
of the gravitational fluctuations $h_{\mu\nu}$. 
After setting $\varphi^A = 0$, the linearized equations of motion for (\ref{Eq. of motion 1})
read
\begin{eqnarray}
&{}& \frac{1}{2} V(H_*) (\partial^\nu h_{\mu\nu} - \frac{1}{2} \partial_\mu h )
+ \frac{\partial^2 V(H_*)}{\partial H^{\mu\nu} \partial H^{\rho\sigma}} 
\partial^\nu h^{\rho\sigma} = 0,
\nonumber\\
&{}& G_{\mu\nu}
= \frac{1}{2} V(H_*) ( \frac{1}{2} \eta_{\mu\nu} h - h_{\mu\nu} )
-  \frac{\partial^2 V(H_*)}{\partial H^{\mu\nu} \partial H^{\rho\sigma}} h^{\rho\sigma},
\label{Linear eq.1}
\end{eqnarray}
where we have used (\ref{Condition 1}) and for simplicity the Einstein's tensor 
$G_{\mu\nu} \equiv R_{\mu\nu} - \frac{1}{2} g_{\mu\nu} R$ is not expanded around
the Minkowski metric.

Then, the strategy for finding appropriate potentials is to require that with a suitable 
choice of the potential the linearized equations of motion (\ref{Linear eq.1}) 
reduce to a set of equations
\begin{eqnarray}
\partial^\nu h_{\mu\nu} - \partial_\mu h &=& 0,
\nonumber\\
G_{\mu\nu}
&=& \frac{m^2}{2} ( \eta_{\mu\nu} h - h_{\mu\nu} ),
\label{Massive Gravity}
\end{eqnarray}
which are the same as those of Fierz-Pauli massive gravity \cite{Fierz}.
This requirement is needed for excluding the appearance of the scalar ghost 
in the spectrum.

In order to find general conditions which the potential must satisfy
to become the Fierz-Pauli massive gravity, by taking account of
the symmetry of indices and the fact that because of the Lorentz symmetry
only the flat Minkowski metric is available as second-rank tensor 
in the linear order of the approximation, we can set 
\begin{eqnarray}
\frac{\partial^2 V(H_*)}{\partial H^{\mu\nu} \partial H^{\rho\sigma}} 
= a_1 \eta_{\mu\nu} \eta_{\rho\sigma}
+ a_2 \eta_{\mu(\rho} \eta_{\sigma)\nu},
\label{Potential 1}
\end{eqnarray}
where $a_1, a_2$ are some constants and we have defined 
$\eta_{\mu(\rho} \eta_{\sigma)\nu} \equiv \frac{1}{2}
( \eta_{\mu\rho} \eta_{\sigma\nu} + \eta_{\mu\sigma} \eta_{\rho\nu} )$.

Substituting Eq. (\ref{Potential 1}) into Eq. (\ref{Linear eq.1}), we obtain
\begin{eqnarray}
&{}& [ \frac{1}{2} V(H_*) + a_2 ] \partial^\nu h_{\mu\nu} 
- [ \frac{1}{4} V(H_*) - a_1 ] \partial_\mu h = 0,
\nonumber\\
&{}& G_{\mu\nu}
= [ \frac{1}{4} V(H_*) - a_1 ] \eta_{\mu\nu} h 
- [ \frac{1}{2} V(H_*) + a_2 ] h_{\mu\nu}.
\label{Linear eq.1-1}
\end{eqnarray}
Then, comparing Eq. (\ref{Linear eq.1-1}) with the Fierz-Pauli massive
gravity (\ref{Massive Gravity}), it turns out that the constants $a_1, a_2$
must satisfy the relations
\begin{eqnarray}
a_1 &=& \frac{1}{4} V(H_*) - \frac{m^2}{2},
\nonumber\\
a_2 &=& - \frac{1}{2} V(H_*) + \frac{m^2}{2}.
\label{Constants 1}
\end{eqnarray}
Consequently, we have a general solution for the potential at $H^{AB}_* =
\eta^{AB}$ which is given by
\begin{eqnarray}
\frac{\partial^2 V(H_*)}{\partial H^{\mu\nu} \partial H^{\rho\sigma}} 
&=& [ \frac{1}{4} V(H_*) - \frac{m^2}{2} ] \eta_{\mu\nu} \eta_{\rho\sigma}
+ [ - \frac{1}{2} V(H_*) + \frac{m^2}{2} ] \eta_{\mu(\rho} \eta_{\sigma)\nu}
\nonumber\\
&=& V(H_*) ( \frac{1}{4} \eta_{\mu\nu} \eta_{\rho\sigma} 
- \frac{1}{2} \eta_{\mu(\rho} \eta_{\sigma)\nu} )
- \frac{m^2}{2} ( \eta_{\mu\nu} \eta_{\rho\sigma} 
- \eta_{\mu(\rho} \eta_{\sigma)\nu} ).
\label{Potential 2}
\end{eqnarray}

To close this section, let us note that the cosmological constant is
defined as $\Lambda \equiv V(H^{AB} = 0)$. Moreover, it is of importance 
to note that in the model of massive gravity under consideration,
although diffeomorphisms are spontaneously broken via condensation of
scalar fields, the Poincare symmetry is never broken. This is a nontrivial
statement since there is in general a possibility such that in the Higgs phase 
of gravitation condensates of scalar fields could break Lorentz and/or 
translational symmetries.

\section{Alternative model based on $g_{\mu\nu}$ and $Y_{\mu\nu}$}

In this section, we wish to construct an alternative general model of Higgs 
mechanism for gravitons based on the usual metric tensor $g_{\mu\nu}$ 
and the induced curved metric $Y_{\mu\nu}$ in a general $D$-dimensional space-time.
We shall proceed in an almost identical fashion to the model treated in
the previous section.

Let us start with the following action:
\begin{eqnarray}
S = \frac{1}{16 \pi G} \int d^D x \sqrt{-g} [ R - V(g_{\mu\nu}, Y_{\mu\nu}) ].
\label{Model 2}
\end{eqnarray}
Here the induced curved metric $Y_{\mu\nu}$ is defined as
\begin{eqnarray}
Y_{\mu\nu} = \eta_{AB} \nabla_\mu \phi^A \nabla_\nu \phi^B,
\label{Y}
\end{eqnarray}
and $V$ is a generic function of both $g_{\mu\nu}$ and $Y_{\mu\nu}$.

{}From the action (\ref{Model 2}), one can easily derive the equations of motion:
\begin{eqnarray}
\nabla_\mu ( \frac{\partial V}{\partial Y_{\mu\nu}} \nabla_\nu \phi^A ) &=& 0,
\nonumber\\
G_{\mu\nu} &=& - \frac{1}{2} g_{\mu\nu} V - g_{\mu\alpha} g_{\nu\beta}
\frac{\partial V}{\partial g_{\alpha\beta}}.
\label{Eq. of motion 2}
\end{eqnarray}

This time, the requirement of the presence of the vacuum solution (\ref{Vacuum})
imposes the following constraint on the potential $V$:
\begin{eqnarray}
\frac{\partial V(g_*, Y_*)}{\partial g_{\mu\nu}} 
= - \frac{1}{2} \eta^{\mu\nu} V(g_*, Y_*),
\label{Condition 2}
\end{eqnarray}
where we have defined $g_{*\mu\nu} = Y_{*\mu\nu} = \eta_{\mu\nu}$. 

As before, after expanding the fields around the vacuum (\ref{Vacuum}) up to
quadratic order and taking the gauge conditions $\varphi^A = 0$, the linearized 
equations of motion for (\ref{Eq. of motion 2}) take the form
\begin{eqnarray}
&{}& \frac{1}{2} \frac{\partial V(g_*, Y_*)}{\partial Y_{\mu\nu}} \partial_\mu h
+ \frac{\partial^2 V(g_*, Y_*)}{\partial Y_{\mu\nu} \partial g_{\rho\sigma}} 
\partial_\mu h_{\rho\sigma} = 0,
\nonumber\\
&{}& G_{\mu\nu}
= \frac{1}{2} V(g_*, Y_*) ( h_{\mu\nu} + \frac{1}{2} \eta_{\mu\nu} h )
-  \eta_{\mu\alpha} \eta_{\nu\beta} \frac{\partial^2 V(g_*, Y_*)}
{\partial g_{\alpha\beta} \partial g_{\rho\sigma}} h_{\rho\sigma},
\label{Linear eq.2}
\end{eqnarray}
where we have used (\ref{Condition 2}).

In order that the linearized equations of motion (\ref{Linear eq.2})
coincide with the Fierz-Pauli massive gravity (\ref{Massive Gravity}),
the potential $V$ must satisfy some contraints at $g_{*\mu\nu} = Y_{*\mu\nu} 
= \eta_{\mu\nu}$. To find them, we assume the following relations:
\begin{eqnarray}
\frac{\partial V(g_*, Y_*)}{\partial Y_{\mu\nu}} &=& b_1 \eta^{\mu\nu},
\nonumber\\
\frac{\partial^2 V(g_*, Y_*)}{\partial Y_{\mu\nu} \partial g_{\rho\sigma}}
&=& b_2 \eta^{\mu\nu} \eta^{\rho\sigma} + b_3 \eta^{\mu(\rho} \eta^{\sigma)\nu},
\nonumber\\
\frac{\partial^2 V(g_*, Y_*)}{\partial g_{\mu\nu} \partial g_{\rho\sigma}}
&=& b_4 \eta^{\mu\nu} \eta^{\rho\sigma} + b_5 \eta^{\mu(\rho} \eta^{\sigma)\nu},
\label{Potential 3}
\end{eqnarray}
where $b_1, \cdots, b_5$ are some constants.

Then, using Eq's. (\ref{Massive Gravity}), (\ref{Linear eq.2}) and (\ref{Potential 3}), 
one is led to the conditions on the constants $b_1, \cdots, b_5$
\begin{eqnarray}
b_3 &=& - ( \frac{1}{2} b_1 + b_2 ) \ne 0,
\nonumber\\
b_4 &=& \frac{1}{4} V(g_*, Y_*) - \frac{m^2}{2},
\nonumber\\
b_5 &=& \frac{1}{2} V(g_*, Y_*) + \frac{m^2}{2}.
\label{Constants 1-1}
\end{eqnarray}
Hence, together with Eq. (\ref{Condition 2}), we have a general solution for 
the potential $V$
\begin{eqnarray}
\frac{\partial V(g_*, Y_*)}{\partial g_{\mu\nu}} 
&=& - \frac{1}{2} V(g_*, Y_*) \eta^{\mu\nu},
\nonumber\\
\frac{\partial V(g_*, Y_*)}{\partial Y_{\mu\nu}} 
&=& b_1 \eta_{\mu\nu},
\nonumber\\
\frac{\partial^2 V(g_*, Y_*)}{\partial Y_{\mu\nu} \partial g_{\rho\sigma}}
&=& b_2 \eta^{\mu\nu} \eta^{\rho\sigma} -  ( \frac{1}{2} b_1 + b_2 )
\eta^{\mu(\rho} \eta^{\sigma)\nu},
\nonumber\\
\frac{\partial^2 V(g_*, Y_*)}{\partial g_{\mu\nu} \partial g_{\rho\sigma}}
&=& [ \frac{1}{4} V(g_*, Y_*) - \frac{m^2}{2} ] \eta^{\mu\nu} \eta^{\rho\sigma} 
+ [ \frac{1}{2} V(g_*, Y_*) + \frac{m^2}{2} ] \eta^{\mu(\rho} \eta^{\sigma)\nu},
\label{Potential 4}
\end{eqnarray}
where $\frac{1}{2} b_1 + b_2  \ne 0$. In other words, the model with any 
potential satisfying Eq. (\ref{Potential 4}) gives rise to physically
plausible massive gravity theories.
Let us note again that the cosmological constant is defined as 
$\Lambda \equiv V(Y_{\mu\nu} = 0)$ as well. 

In this section, we have presented an alternative model of the Higgs mechanism
for gravitons based on metrics $g_{\mu\nu}$ and $Y_{\mu\nu}$. At first sight, 
it might appear that this new massive gravity model is different from the model 
made in the previous section. However, this is an illusion. In fact, the two kinds 
of models are essentially equivalent since we have identities such as 
$\eta_{AB} H^{AB} = g^{\mu\nu} Y_{\mu\nu}, H^{AB} H_{AB} = Y_{\mu\nu} Y^{\mu\nu}$
etc. Although the two models are equal to each other, the model based
on $H^{AB}$ is more convenient to handle than that based on $g_{\mu\nu}$ and $Y_{\mu\nu}$
in the sense that Eq's. (\ref{Condition 1}) and (\ref{Potential 2}) are simpler 
than Eq. (\ref{Potential 4}).
Thus, we shall make use of the model based on the induced internal metric $H^{AB}$ 
for presenting examples in the next section.

\section{Concrete models}

We are now ready to present some concrete models of the Higgs mechanism for
gravitons by fixing the form of the potential $V$. Before doing so, let us
recall relevant works done thus far. In Ref. \cite{'t Hooft}, 
't Hooft has advocated an idea of Higgs mechanism for gravitons where 
the massless gravitons '$\it{eat}$' four real scalar fields and consequently
become massive. One serious problem in the 't Hooft model is 
that a scalar field appearing after the SSB is a non-unitary propagating field, 
thereby violating the unitarity. This problem was 
afterward solved by including higher-derivative terms in scalar fields 
and tuning appropriately the cosmological constant to be a negative value 
in Ref. \cite{Kaku2}. 

Recently, Chamseddine and Mukhanov have presented a new Higgs mechanism for
gravitons also by adding a specific combination of higher-derivative terms 
in scalar fields to the Einstein-Hilbert
action \cite{Chamseddine}. One advantage of their model is that we do not have to
restrict the cosmological constant to be negative, namely zero or positive cosmological
constant is also allowed to trigger the gravitational Higgs mechanism. 

Although the model by Chamseddine and Mukhanov is of interest, the potential which they
have found is a bit tricky, higher order in $H^{AB}$ (in fact, sixth order) 
and has a special property $V(H_*) = 0$, so it might be more interesting if we could 
find more natural models with the lower-order terms in $H^{AB}$ which also exhibit 
the Higgs mechanism for gravitons in any value of the cosmological constant. 
In this section, we shall look for such a model.

As the first model, we would like to deal with a model with the property that
the cosmological constant takes any value except in two and four dimensions.
The first model has quadratic terms with respect to the metric $H^{AB}$:
\begin{eqnarray}
V(H^{AB}) = \Lambda + \alpha_1 H + \alpha_2 H_{AB} H^{AB} + \alpha_3 H^2,
\label{Example 1}
\end{eqnarray}
where $H \equiv \eta_{AB} H^{AB}$, $\Lambda$ is the cosmological constant, 
and $\alpha_1, \alpha_2, \alpha_3$ are constants to be determined shortly. 

The condition (\ref{Condition 1}) gives rise to
\begin{eqnarray}
\alpha_1 + 2 \alpha_2 + 2 D \alpha_3 = \frac{1}{2} V(H_*).
\label{Constraint 1}
\end{eqnarray}
Moreover, Eq. (\ref{Potential 2}) produces the relations  
\begin{eqnarray}
\alpha_2 &=& - \frac{1}{4} V(H_*) + \frac{m^2}{4},
\nonumber\\
\alpha_3 &=& \frac{1}{8} V(H_*) - \frac{m^2}{4}.
\label{Constraint 2}
\end{eqnarray}
Together with Eq's. (\ref{Constraint 1}) and (\ref{Constraint 2}), we can
express the remaining $\alpha_1$ by
\begin{eqnarray}
\alpha_1 = ( 1 - \frac{D}{4} ) V(H_*) + \frac{D-1}{2} m^2.
\label{Constraint 3}
\end{eqnarray}
As a result, the potential reads
\begin{eqnarray}
V(H^{AB}) &=& \Lambda + [( 1 - \frac{D}{4} ) V(H_*) + \frac{D-1}{2} m^2] H 
+ [- \frac{1}{4} V(H_*) + \frac{m^2}{4}] H_{AB} H^{AB} 
\nonumber\\
&+& [\frac{1}{8} V(H_*) - \frac{m^2}{4}] H^2.
\label{Result 1}
\end{eqnarray}

Then, with the help of (\ref{Result 1}), putting $H^{AB} = H^{AB}_* = \eta^{AB}$
leads to the value of the cosmological constant in this model
\begin{eqnarray}
\Lambda = \frac{(D-2)(D-4)}{8} V(H_*) - \frac{D(D-1)}{4} m^2.
\label{CC1}
\end{eqnarray}
Depending on the value of $V(H_*)$, the cosmological constant $\Lambda$
takes an arbitrary value. However, note that in particular, 
$\Lambda = - 3 m^2 < 0$ for $D = 2, 4$, so this model of the Higgs mechanism for gravitons 
holds only in case of the negative cosmological constant in our four-dimensional space-time,
which is a unsatisfactory point of this model.

Nevertheless, the model at hand gives us two useful informations. The one information
is that our model includes the model constructed by Kakushadze in a specific case \cite{Kaku2}. 
Indeed, his model corresponds to the case of $\alpha_2 = 0$, that is, 
$V(H_*) = m^2$, and then the cosmological constant is given by 
$\Lambda = - \frac{D^2 + 4 D - 8}{8} m^2$ and is always negative for $D>1$ \cite{Kaku2}.
The other information extracted from our model is that we cannot construct
a plausible massive gravity model by starting with the 't Hooft model.
The 't Hooft model consists of only linear term of $H$ in addition to the constant cosmological
term, so in order to get the 't Hooft model, we have to set both $\alpha_2$ and $\alpha_3$ 
to be zero at the same time. But it is obviously impossible, thus meaning that
the 't Hooft model is not free from the ghost mode.

A problem associated with the first model (\ref{Result 1}) is that we can construct a reasonable massive
gravity model only in case of the negative cosmological constant in four dimensions.
Thus, next let us move on to the second model of the Higgs mechanism for gravitons, which is a slight
generalization of the first model in that the potential now involves cubic terms
in $H^{AB}$, but the second model turns out to be free from the issue of the cosmological constant. 

The potential of the second model is given by
\begin{eqnarray}
V(H^{AB}) &=& \Lambda + \alpha_1 H + \alpha_2 H_{AB} H^{AB} + \alpha_3 H^2
+ \alpha_4 H^3 + \alpha_5 H H_{AB} H^{AB} 
\nonumber\\
&+& \alpha_6 H_{AB} H^{BC} H_C \ ^A,
\label{Example 2}
\end{eqnarray}
where $\alpha_1, \cdots, \alpha_6$ are constants to be determined later. 

As in the first model, the conditions (\ref{Condition 1}) and (\ref{Potential 2}) 
lead to the relations among $\alpha_i (i = 1, \cdots, 6)$, but it turns out that they
only determine $\alpha_3, \alpha_5, \alpha_6$ in terms of remaining $\alpha_1, \alpha_2, 
\alpha_4$ whose result reads
\begin{eqnarray}
\alpha_3 &=& - \frac{1}{D} [ \alpha_1 + \alpha_2 + \frac{D-6}{8} V(H_*) 
- \frac{D-1}{4} m^2 ],
\nonumber\\
\alpha_5 &=& \frac{1}{4} [ \frac{2}{D} ( \alpha_1 + \alpha_2 ) - 6 D \alpha_4
+ \frac{D-3}{2D} V(H_*) - \frac{2D-1}{2D} m^2 ],
\nonumber\\
\alpha_6 &=& \frac{1}{6} [ - \alpha_1 - 3 \alpha_2 + 3 D^2 \alpha_4 
- \frac{D-1}{4} V(H_*) + \frac{2D+1}{4} m^2 ].
\label{Constraint 4}
\end{eqnarray}

Then, it is straightforward to calculate the cosmological constant 
defined as $\Lambda \equiv V(H^{AB} = 0)$ as before, which reads
\begin{eqnarray}
\Lambda = - \frac{D}{3} \alpha_1 + \frac{(D-4)(D-6)}{24} V(H_*) - \frac{D(D-1)}{12} m^2.
\label{CC2}
\end{eqnarray}
In this case, it is remarkable that since $\Lambda = - \frac{4}{3} \alpha_1 - m^2$ for $D = 4$
we can construct a massive gravity model at any value of the cosmological constant
by selecting out the value of $\alpha_1$ in an appropriate way. Hence, owing to the existence
of cubic terms,  this model of the Higgs mechanism for gravitons holds irrespective of  
the signature of the cosmological constant in four dimensions.

As a final model, for completeness, let us comment on a model presented recently by 
Chamseddine and Mukhanov in Ref. \cite{Chamseddine} in the framework of this article.
The potential of their model is of sixth-order in $H^{AB}$ and is explicitly given by
\begin{eqnarray}
V(H^{AB}) = e_1 [ (\frac{1}{D} H)^2 - 1 ]^2
[ \alpha (\frac{1}{D} H)^2 - \beta ] + e_2 H_{AB} H^{AB} + e_3 H^2,
\label{Example 3}
\end{eqnarray}
where $e_1, e_2, e_3, \alpha, \beta$ are constants. 

Again, the conditions (\ref{Condition 1}) and (\ref{Potential 2}) determine
the constants $e_1, e_2, e_3$ as
\begin{eqnarray}
e_1 &=& \frac{D}{8 (\alpha - \beta)} [ \frac{D-4}{4} V(H_*) - \frac{D-1}{2} m^2 ],
\nonumber\\
e_2 &=& - \frac{1}{4} V(H_*) + \frac{1}{4} m^2,
\nonumber\\
e_3 &=& \frac{1}{2D} [ V(H_*) - \frac{1}{2} m^2 ],
\label{Constraint 5}
\end{eqnarray}
where $\alpha - \beta \ne 0$.

Then, the cosmological constant given by $\Lambda = - e_1 \beta$ reads
\begin{eqnarray}
\Lambda = - \frac{\beta D}{8 (\alpha - \beta)} [ \frac{D-4}{4} V(H_*) - 
\frac{D-1}{2} m^2 ].
\label{CC3}
\end{eqnarray}
In particular, for $D = 4$, we have
\begin{eqnarray}
\Lambda = \frac{3 \beta}{4 (\alpha - \beta)} m^2,
\label{CC4}
\end{eqnarray}
which can certainly take any value depending on the values of $\alpha$ and $\beta$.

\section{Conclusion and discussions}

In this article, we have constructed two kinds of model exhibiting Higgs mechanism for 
gravitons where the one class is based on a potential of the induced internal metric $H^{AB}$, 
and the other class involves a potential of the usual metric tensor $g_{\mu\nu}$ and 
the induced curved metric $Y_{\mu\nu}$. Even if they appear to be different models at
first sight, they are in fact equivalent because of identities among metrics.
Furthermore, using the former model, we have explicitly presented a massive gravity model
holding at any value of the cosmological constant.

Incidentally, our formalism reminds us of bimetric theory of gravity which
was made by Rosen \cite{Rosen} long ago. Actually we have the induced metric
$H^{AB}$ or $Y_{\mu\nu}$ made out of scalar fields as well as the metric tensor $g_{\mu\nu}$,
and the induced metric plays a role of the order parameter in spontaneous symmetry breakdown.
Namely, the induced metric $H^{AB}$ or $Y_{\mu\nu}$ constructed out of scalar fields
condenses to produce mass for the metric tensor $g_{\mu\nu}$. Related to this observation,
our formalism might be concerned with strong gravity by Salam and Strathdee \cite{Salam}.

Moreover, it is useful to recall that at present we have an alternative mechanism to give 
mass to massless gravitons only in three dimensions. In three dimensions, 
the new massive gravity (NMG) provides us non-linear, parity invariant, and
generally covariant massive gravitons by adding a conformal combination of 
curvature squared terms to the Einstein-Hilbert term \cite{Bergshoeff}.
This NMG turns out to be unitary at least at the tree level \cite{Nakasone1}
and super-renormalizable \cite{Oda3}.
Then, it is of interest to examine a relation between the NMG and the Higgs mechanism
for gravitons. Indeed, for instance, we have already shown that the NMG cannot
coexist with the Fierz-Pauli term \cite{Nakasone2}, so it is an interesting question 
to investigate whether the NMG could coexist with the Higgs mechanism for gravitons at hand.

What remains to be seen is an explicit calculation to demonstrate how the
massless limit is attained in the framework of the Higgs mechanism for
gravitons.  The other interesting problem is to examine whether the present models
are really consistent models or not. Recall that the classical treatment 
by Fierz and Pauli \cite{Fierz}, which is perfectly satisfactory for free
fields though, meets with difficulties when interactions are switched on. 
In the models at hand, the graviton mass is generated by
the dynamical mechanism, i.e., Higgs mechanism, so there would be a possibility
to escape this problem. Anyway we would like to report these problems in future.
   
Recall that in case of massive Yang-Mills theory the Higgs mechanism made it
possible to not only achieve a well-defined massless continuity but also
allow us to have a renormalizable theory of massive vector gauge fields
which stay in the weak coupling regime. On the other hand, in case of gravity
the Higgs mechanism would only allow a smooth transition from massless gravitons
to massive gravitons. Since the Einstein-Hilbert action is known to be
unrenormalizable, the corresponding massive gravity theory is also unrenormalizable
and it might behave much more badly in the ultraviolet region since we have
added the higher-derivative coupling terms of scalar fields in the potential.
In any case, quantum corrections to massive gravity theory would occur at
some cut-off scale, so the corrections could be ignored at energies much
lower than the cut-off scale.   

A detailed study of the massive gravity has been done in the paper by Boulware and
Deser \cite{Boulware}. While the theory of massive gravity is well-defined at the 
linear level, it becomes to have extra sixth trace mode, which is in essence a ghost,
in addition to five degrees of freedom at the non-linear level in four dimensions.
We can write down several questions which were already raised up about fourty
years ago \cite{Boulware}.
\begin{itemize}
\item The massless limit does exist and agree with general relativity?
\item The energy is bounded from below?
\item The flat space-time is a local stable equilibrium state?
\item When interactions are switched on, does the trace mode appear, thereby
breaking the unitarity?
\end{itemize}   

We believe that our models of the Higgs mechanism for gravitons could provide 
a resolution for the above-mentioned questions. A detailed analysis will
be reported in a separate publication. We wish to close this article by
confessing a philosophy behind this study: The concept of spontaneous symmetry breaking
prevails and have provided a considerable influence on the evolution of particle
physics and the condensed matter physics, so why the most universal interaction,
gravitation, does not adopt such a beautiful concept in the theory?!

\section*{Acknowledgments}

This work is supported in part by the Grant-in-Aid for Scientific Research (C)
No. 22540287 from the Japan Ministry of Education, Culture, Sports,
Science and Technology.

\vs 1   


\begin{thebibliography}{99}

\bibitem{Fierz}
M. Fierz and W. Pauli, {"On Relativistic Wave Equations for Particles of Arbitrary 
Spin in an Electromagnetic Field", Proc. Roy. Soc. Lond. 
{\bf A173} (1939) 211.}

\bibitem{Vainshtein}
A. Vainshtein, {"Massive Gravity", Surveys in High Energy Physics 
{\bf 20} (2006) 5. }

\bibitem{Veltman}
H. van Dam and M. Veltman, {"Massive and Massless Yang-Mills and
Gravitational Fields", \NP {\bf B22} (1970) 397.}

\bibitem{Zakharov}
V.I. Zakharov, {"Linearized Gravitation Theory and the Graviton Mass", 
JETP Lett. {\bf 12} (1970) 312.}

\bibitem{Percacci}
R. Percacci, {"The Higgs Phenomenon in Quantum Gravity",
\NP {\bf B353} (1991) 271, arXiv:0712.3545 [hep-th]};
C. Omero and R. Percacci, {"Generalized Nonlinear Sigma Models in Curved Space
and Spontaneous Compactification", \NP {\bf B165} (1980) 351.}

\bibitem{Kaku1}
Z. Kakushadze and P. Langfelder, {"Gravitational Higgs Mechanism", 
\MPL {\bf A15} (2000) 2265, arXiv:hep-th/0011245.}

\bibitem{Porrati}
M. Porrati, {"Higgs Phenomenon for 4-D Gravity in Anti de Sitter Space",
JHEP {\bf 0204} (2002) 058, arXiv:hep-th/0112166.}

\bibitem{Leclerc}
M. Leclerc, {"The Higgs Sector of Gravitational Gauge Theories", \AP {\bf 321} 
(2006) 708, arXiv:gr-qc/0502005.}

\bibitem{Georgi}
N. Arkani-Hamed, H. Georgi and M.D. Schwartz, {"Effective Field Theory for Massive 
Gravitons and Gravity in Theory Space", \AP {\bf 305} (2003) 96, 
arXiv:hep-th/0210184.}

\bibitem{'t Hooft}
G. 't Hooft, {"Unitarity in the Brout-Englert-Higgs Mechanism for Gravity",
arXiv:0708.3184 [hep-th].}

\bibitem{Kaku2}
Z. Kakushadze, {"Gravitational Higgs Mechanism and Massive Gravity",
\IJMP {\bf A23} (2008) 1581, arXiv:0709.1673 [hep-th]; "Massive Gravity in Minkowski 
Space via Gravitational Higgs Mechanism", \PR {\bf D77} (2008) 024001, 
arXiv:0710.1061 [hep-th].}

\bibitem{Maeno}
M. Maeno and I. Oda, {"Classical Solutions of Ghost Condensation Models", 
\MPL {\bf B22} (2009) 3025, arXiv:0801.0827 [hep-th]; "Massive Gravity in Curved 
Cosmological Backgrounds", \IJMP {\bf A24} (2009) 81-100, arXiv:0808.1394 [hep-th].} 

\bibitem{Rubakov2}
V.A. Rubakov and P.G. Tinyakov, {"Infrared-modified Gravities and Massive
Gravitons", arXiv:0802.4379 [hep-th].}

\bibitem{Riess}
A.G. Riess et al., {"Observational Evidence from Supernovae for an Accelerating 
Universe and a Cosmological Constant", Astron. J. {\bf 116} (1998) 1009,
arXiv:astro-ph/9805201.}

\bibitem{Perlmutter}
S. Perlmutter et al., {"Measurements of Omega and Lambda from 42 High Redshift 
Supernovae", Astron. J. {\bf 517} (1999) 565,
arXiv:astro-ph/9812133.}

\bibitem{WMAP}
WMAP Collaboration, D.N. Spergel et al., {"Wilkinson Microwave Anisotropy Probe (WMAP) 
Three Year Results: Implications for Cosmology", Astron. J. Suppl. {\bf 170} 
(2007) 377, arXiv:astro-ph/0603449.}

\bibitem{Oda}
I. Oda, {"Gravitational Higgs Mechanism with a Topological Term", 
Adv. Studies Theor. Phys. {\bf 2} (2008) 261, arXiv:0709.2419 [hep-th].}

\bibitem{Oda1}
I. Oda, {"Strings from Black Hole", \IJMP {\bf D1} (1992) 355; K. Akama and I. Oda,
"Topological Pregauge Pregeometry", \PL {\bf 259} (1991) 431;
K. Akama and I. Oda, "BRST Quantization of Pregeometry and Topological Pregeometry",
\NP {\bf B397} (1993) 727.}

\bibitem{Demir}
D.A. Demir and N.K. Pak, {"General Tensor Lagrangian from Gravitational Higgs 
Mechanism", \CQG {\bf 26} (2009) 105018, arXiv:0904.0089 [hep-th]. }

\bibitem{Chamseddine}
A.H. Chamseddine and V. Mukhanov, {"Higgs for Gravitons: Simple and Elegant Solution", 
arXiv:1002.3877 [hep-th].}

\bibitem{Oda2}
I. Oda, {"Higgs Mechanism for Gravitons", arXiv:1003.1437 [hep-th].}

\bibitem{MTW}
C.W. Misner, K.S. Thorne and J.A. Wheeler, {"Gravitation", W H Freeman and 
Co (Sd), 1973.}

\bibitem{Rosen}
N. Rosen, {"General Relativity and Flat Space. I", \PR {\bf 57} 
(1940) 147; "General Relativity and Flat Space. II", \PR {\bf 57} 
(1940) 150.}

\bibitem{Salam}
A. Salam and J. Strathdee, {"Class of Solutions for the Strong-gravity Equations", 
\PR {\bf 16} (1977) 2668.}

\bibitem{Bergshoeff}
E.A. Bergshoeff, O. Hohm and P.A. Townsend, {"Massive Gravity in Three Dimensions",
\PRL {\bf 102} (2009) 201301,
arXiv:0901.1766 [hep-th].}

\bibitem{Nakasone1}
M. Nakasone and I. Oda, {"On Unitarity of Massive Gravity in Three Dimensions",
\PTP {\bf 121} (2009) 1389,
arXiv:0902.3531 [hep-th].} 

\bibitem{Oda3}
I. Oda, {"Renormalizability of Massive Gravity in Three Dimensions", 
JHEP {\bf 0905} (2009) 064, 
arXiv:0904.2833 [hep-th].} 

\bibitem{Nakasone2}
M. Nakasone and I. Oda, {"Massive Gravity with Mass Term in Three Dimensions",
\PR {\bf D79} (2009) 104012,
arXiv:0903.1459 [hep-th].} 

\bibitem{Boulware}
D.G. Boulware and S. Deser, {"Can Gravitation Have a Finite Range?", \PR {\bf D6} 
(1972) 3368.}



\end{thebibliography}
\end{document}